# WISHBONE BUS ARCHITECTURE – A SURVEY AND COMPARISON

Mohandeep Sharma[1] and Dilip Kumar[2]

[1]Department of VLSI Design, Center for Development of Advanced Computing, Mohali, India
mdsharma007@gmail.com
[2]ACS - Division, Center for Development of Advanced Computing, Mohali, India
dilip.k78@gmail.com

*ABSTRACT*

*The performance of an on-chip interconnection architecture used for communication between IP cores depends on the efficiency of its bus architecture. Any bus architecture having advantages of faster bus clock speed, extra data transfer cycle, improved bus width and throughput is highly desirable for a low cost, reduced time-to-market and efficient System-on-Chip (SoC). This paper presents a survey of WISHBONE bus architecture and its comparison with three other on-chip bus architectures viz. Advanced Microcontroller Bus Architecture (AMBA) by ARM, CoreConnect by IBM and Avalon by Altera. The WISHBONE Bus Architecture by Silicore Corporation appears to be gaining an upper edge over the other three bus architecture types because of its special performance parameters like the use of flexible arbitration scheme and additional data transfer cycle (Read-Modify-Write cycle). Moreover, its IP Cores are available free for use requiring neither any registration nor any agreement or license.*

## 1. INTRODUCTION

The introduction and advancement of multimillion-gate chips technology with new levels of integration in the form of the system-on-chip (SoC) design has brought a revolution in the modern electronics industry. With the evolution of shrinking process technologies and increasing design sizes [1], manufacturers are integrating increasing numbers of components on a chip. Researchers and producers, with every new day, intend to include programmable components like the general-purpose processors cores, digital signal processor cores, or application-specific intellectual property (IP) cores, on-chip memory, and other application-specific circuits on a SoC turning it into a highly complex multimillion transistor integrated circuits. Bernini rightly concludes that a SoC these days has turned into an IC that implements most or all the functions of a complete electronic system [2].

However, these developments are not free from problems and challenges. In addition to the testing of performances of topologies in terms of primary parameters like area, delay and power dissipation [3]**, t**he biggest challenge, for designers, is the implementation of the on-chip interconnections [4] while using buses in the SoC design. With ever increasing emphasis on the application of the design re-use, the situation for the designers has become more complex and the task of bus interfacing and interconnections has turned into a highly complicated affair. It is a well known fact that the balance of computation and communication in any application is quite a significant determinant of delivered performance. Therefore, the designing and integration of IP cores, having different interfaces and communication protocols, as constituents of SoCs not only assume a significant proportion but also turn out to be a major challenge for the designers. One of the possible solutions lies in the use of standard internal connection bus for interconnecting design components in SoC application modules so as to make it quite convenient for the present and the future re-use.

The present VLSI design scenario under the constraints of rapidly increasing operation frequencies and ever growing chip size brings the designers face to face with the problem of the on-chip bus organized communication architecture in SoC technology. The system performance is dependent on the CPU speed as well as on the efficient bus architecture [5] which along with arbitration helps in maximizing the performance of the system by reducing contention. Being aware of the fact that the very efficiency of the bus architecture is generally responsible for the performance of the SoC design, the manufacturers like the IBM, ARM, ALTERA, SILICORE etc. tried successfully solving this problem of organized communication architecture by developing standards of on-chip bus structures and making the same publicly available.

This paper attempts mainly to present a survey of the WISHBONE bus architecture developed by the SILICORE [6], and making a comparison of its selective features with that of some other popular on-chip standardized bus architectures such as AMBA [7], CoreConnect [8], Avalon [9], etc. The paper endeavors to survey and review the features like the bus topologies, communication protocols, arbitration methods, bus-widths, and types of data transfers in these SoC organized communication architectures.

The organization scheme of the paper is as follows: Section 2 presents the background and description of the WISHBONE bus architecture including WISHBONE interfaces, protocols and signal mapping. Section 3 focuses on the overview of three SoC bus architecture types namely, AMBA, CoreConnect and Avalon. Section 4 attempts a comparison of the four Soc Bus Architecture types and includes a table of comparison. Finally, the conclusion of this paper is presented in Section 5.

## *2.* ON-CHIP WISHBONE BUS ARCHITECTURE

### *2.1.* Background

The WISHBONE specification document [10] defines the WISHBONE bus as the System-on-Chip (SoC) architecture which is a portable interface for use with semiconductor IP cores. It is intended to be used as an internal bus for SoC applications with the aim of alleviating SoC integration problems by fostering design reuse. This objective is achieved by creating a common interface between IP cores [11]. It improves the portability, reliability of the system, and results in faster time-to-market for the end user [12]. However, the cores can be integrated more quickly and easily by the end user if a standard interconnection scheme is adopted. The WISHBONE bus helps the end user to accomplish all these objectives at one platform. Intended as a general purpose interface, WISHBONE bus defines a standard set of signals and bus cycles to be used between IP core modules making no attempt to regulate the application specific functions of the IP core. On-Chip design of WISHBONE Bus Architecture addresses the following three issues:

### 2.1.1 Definition of WBA Interface

The issue of the physical structure of the Wishbone Bus Architecture is defined under the WBA Interface. There are two types of interfaces under the WISHBONE bus. These are called MASTER and SLAVE interfaces. Cores that help to generate bus cycles are identified as MASTER interfaces and the cores capable of receiving bus cycles are designated as SLAVE interfaces. As these interfaces range from single to complex architectures, there are a number of ways to connect between the MASTER and the SLAVE interfaces in a WISHBONE bus. These include Point-to-point interconnection, Data flow interconnection, Shared bus interconnection and Crossbar switch interconnection. All these different ways of interface interconnections have been elaborated under sub section 2.2 below.

### 2.1.2 Selection and configuration of WISHBONE Bus Protocols

WISHBONE Bus protocols specify the manner in which transactions occur. These protocols include the implementation of arbitration mechanism in centralized or distributed bus arbiters. The issue of bus contention during the selection and configuration of WISHBONE bus protocol is settled or decided with the help of Handshaking protocol; deployment of different arbitration schemes such as Round Robin, TDMA, CDMA, Static Priority, Token Passing etc. All these strategies are application specific in WISHBONE Bus. These issues find a detailed elaboration under sub section 2.3 of this paper.

### 2.1.3 Signal Mapping

Signal Mapping under WISHBONE bus is a process of associating Master and Slave devices in the bus Architecture. It also includes Bus Cycles. Although WISHBONE allows combining of all its signals between the MASTER and SLAVE interfaces yet each can do it at its own expense. WISHBONE signals have been grouped into three categories: Common Signals, Data Signals, and Bus Cycle Signals [13]. The issues of WISHBONE signal types and WISHBONE bus cycles have been dealt with relevant details in sub section 2.4 of this paper.

## 2.2 Interfaces

Bus interfacing involves an electronic circuit that is responsible for driving or receiving data or power from a bus. So far as the interfacing in the WISHBONE bus is concerned, the on-chip WISHBONE bus architectures [14] can be classified as: Point-to-point interconnection, Shared bus interconnection, Crossbar switch interconnection, Data flow interconnection, and Off-chip interconnection. As the implementation of the Off-chip interconnection, fits generally into one of the other four basic types, therefore it is not included under a separate heading for discussion in this paper. The other four basic types of WISHBONE interconnections are:

### 2.2.1 Point to Point Interconnection

A Point-to-Point interconnection supports direct connection of two participants that transfer data according to some handshake protocol. It implies that a single master has a direct connection to a single slave. This is the simplest way of connecting two IP cores and the traffic is controlled by the handshaking signals. As the Point-to-point INTERCON only supports connection of a single master interface and a single slave interface, its limitations do not make it suitable for SoC multi-device inter-connection.

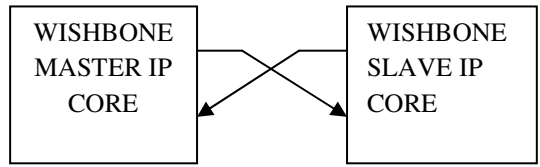

Figure – 1 Point to Point Interconnection

### 2.2.2 Shared Bus Interconnection

In a Shared Bus interconnection many masters and slaves share the bus with each other. However, only one master at a time can use the bus, and the other masters have to wait for their turn. An arbiter controlling the bus decides which master may use it at a particular moment. As a Shared bus INTERCON supports a single channel connection allowing only one master to initiate a bus cycle to a target slave through connected channel at a time, the data transfer rate of the shared bus INTERCON also turns out to be of limited nature.

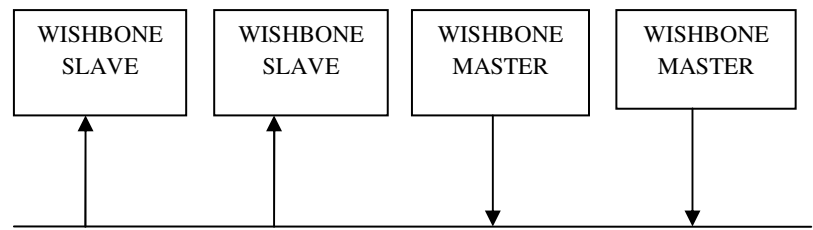

Figure – 2 Shared Bus Interconnection

### 2.2.3 Crossbar Switch Interconnection

A Crossbar Switch Interconnection WISHBONE supported topology can be used in multi core SoCs where more than one master can simultaneously access several slaves. There are multiple ways for data to be transferred between masters and slaves in a crossbar switch interconnection. Therefore two or more masters can communicate with slaves at the same time, as long as it isn't the same slaves. As compared to a shared bus, it leads to a higher data transfer rate. In this type of interconnection, there is always an arbiter to control the bus. Arbiter decides which master may communicate with which slave. As the use of a cross-bar interconnect system in WISHBONE enables multiple masters to communicate, concurrently, with multiple slaves, these interconnect has a limited scalability due to a centralized resource. The square of connected component numbers increases complexity.

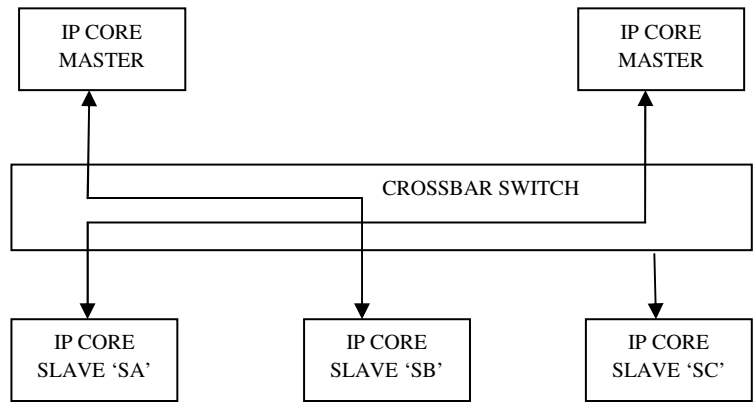

Figure – 3 Crossbar Switch Interconnection

### 2.2.4 Data Flow Interconnection

A Data Flow interconnection works in such a way as the data is processed in a sequential manner. Every IP core in the data flow architecture has both a MASTER and a SLAVE interface. An IP Core acts as a master to the next IP Core in the sequential chain and as a slave to the IP Core prior to it. Data flows from core-to-core. Sometimes this process is called pipelining. The data flow architecture also exploits parallelism, thereby speeding up execution time. The traffic is controlled by the handshaking signals. A dataflow interconnection topology supported by WISHBONE can be useful for linear systolic array architectures used in implementation of DSP algorithms [15].

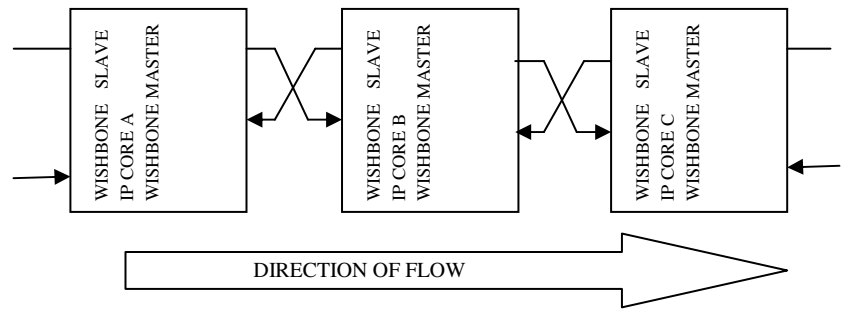

Figure – 4 Data Flow Interconnection

### 2.3. *WISHBONE PROTOCOLS*

All Communication Protocols deal with different types of algorithms which are used for determining access to different interconnections [16]. A brief survey pertaining to the main protocol strategies and the main features of the existing protocols with special reference to the particular bus type making use of the same is presented as under:

### 2.3.1 Static – Priority

The Static-Priority protocol is used in shared bus interconnection architectures. It employs an arbitration technique wherein centralized arbiter examines requests accumulated from each master and grants access to the master of the highest priority. The use of this arbitration technique in WISHBONE protocol is application specific while in Avalon bus protocol [9] it is on the slave side or distributed arbitration. Like WISHBONE bus, AMBA bus [7] also deploys Static-Priority protocol which is application specific except in the case of the APB or the Advanced Peripheral Bus.

### 2.3.2 Round Robin

This protocol deploys an arbitration mechanism which makes use of statically assigned arbitration numbers for resolving the conflict during the process of arbitration [17]. Under the Round Robin mechanism every master shares the bus with each slave device. Appropriately, this protocol is used in crossbar- bus interconnection. The use of this protocol is application specific.

### 2.3.3 Time Division Multiple Access (TDMA)

The arbitration Mechanism under TDMA is based on a timing wheel with each slot statically reserved for a unique master. The problem of wasted slots under this protocol is tackled by using special techniques. AMBA, AVALON and WISHBONE deploy this protocol.

### 2.3.4 Lottery

Under this protocol, the request for ownership of shared communication resources from one or more masters is accumulated with the help of a centralized lottery manager. Each and every one of these masters is, statically or dynamically, assigned a number of "lottery tickets" [18].

### 2.4 Signal Mapping

This section presents a brief overview of signals of the WISHBONE bus and the bus cycles.

### 2.4.1 WISHBONE Signals

The WISHBONE interface consists of signals. These are master signals, slave signals and signals that are common to both masters and slaves. All WISHBONE signals use active high logic. Table No.1 helps in understanding the process how a master is connected to a slave. It also highlights whether the signals are optional or not for a WISHBONE interface. As evident from Table-1, it is not necessary for an IP Core to implement all signals to be WISHBONE compatible. For example the signals for handling error detection are optional.

Table – 1 WISHBONE Signals

| Signal | Name | Optional | Description |
|---|---|---|---|
| Acknowledged Out | ACK_O | No | Acknowledged signal from the Master to Slave |
| Address Out/In | ADR_O/I() | No | Address Array |
| Clock In | CLK_I | No | System Clock for Wishbone Interface. |
| Error In/Out | ERR_I/O | Yes | Indicates an abnormal cycle termination occurred. |
| Data In/Out | DAT_I/O | No | Data input/output array, used to send/receive data. |
| Write Enable In | WE_I | No | Read or Write Signals. If asserted it is Write signal otherwise Read Signal. |
| Strobe In | STB_I | No | Indicates that the slave is selected The slave asserts either ACK_O, ERR_O |
| Strobe Out | STB_O | No | Handshaking Signal. |
| Address tag Out | TGA_O() | Yes | Contains Information about the address array. |
| Cycle tag type Out | TGC_O() | Yes | Contains Information about the transfer cycle. |
| Write Enable Out | WE_O | No | Shows if the transfer cycle is a Read or Write cycle. |
| Reset | RST_I | No | Reset signal |

### 2.4.2  WISHBONE Bus Cycles

WISHBONE supports three types of bus cycles; Single Read/Write cycles, Block Read/Write cycles and Read-Modify-Write cycles.

### 2.4.2.1  Single Read/Write Cycle

A single read/write cycle means that only one data transfer is made each time. For example when a master wants to make a single read operation it presents a valid address on its address out port. Then it negates the write enable signal to show that a read operation is to be done. After that it asserts its cyclic out and strobe out signals to tell the slave that the transfer is ready to start. When the slave has noticed the assertion of the strobe and cyclic signals it places the right data on the data out port and asserts its acknowledged out signal. At the next clock edge the master will read the data and pull down its strobe out and cyclic out signal, which leads to that the slave negates its acknowledged out signal and thereby the transmission is complete.

### 2.4.2.2 Block Read/Write Cycle

Block Read/Write cycles are used when a master wants to read or write multiple data arrays and works in a similar way as the single read/write cycle. The main difference is that the negation of the cyclic signal from the master does not occur until all data is transferred, instead the strobe and acknowledge signals control the flow of data arrays between the master and the slave. The master can put in wait states by pulling down its strobe signal, whereas the slave can put in wait states by pulling down it's acknowledge signal.

### 2.4.2.3 Read-Modify-Write (RMW) Cycle

RMW cycle is used to avoid the possibility of two or more masters gaining access to the same slave. Such a possibility may occur in systems having multiple processors that share memories. It becomes important to ensure that they don't access the same memory at the same time. To prevent such a happening, a slave under use must be blocked. This is often done by assertion of a semaphore bit. If a master reads that a semaphore bit is asserted it knows that the slave is accessed by another master. The master has to use a RMW cycle in such a situation. First the master reads the semaphore bit and if it is cleared the master will assert it by writing something to it during the same transfer cycle. If the same procedure would be done by using single read and write operations another master could try to access the slave between the read and write operation, leading to that both masters would get access to the slave at the same time. In other words a RMW cycle gives a master the opportunity to both do a read and a write operation before any other master may use the bus and thereby avoiding a system crash.

## *3.* SOC BUSES OVERVIEW

As an overview of the WISHBONE bus has been attempted in the previous section, therefore the focus in this section is on presenting an overview of only three SoC buses -- AMBA, CoreConnect and Avalon. Due to space limitation, the strategy used here in this section is to rely on describing the selective and more distinctive features of every one of these buses.

### 3.1 AMBA

AMBA is an abbreviation of the *Advanced Microcontroller Bus Architecture*. This particular bus standard has been devised by ARM [7] to support on-chip communications among the processor cores manufactured by this particular company. AMBA is, nowadays, one of the leading on-chip bus systems used in high performance SoC design. The most important issue in a SoC these days is not only the housing of components or blocks but also the way these are interconnected. AMBA provides an efficient solution for the blocks to interface with each other [19]. AMBA (shown in Fig. 5) is organized hierarchically into two bus segments --- System- bus and Peripheral-bus, which are mutually connected through a bridge that serves to buffer data and operations between these segments. Specifications issued along with AMBA define the Standard bus protocols used for connecting on-chip components generalized for different SoC structures, and independent of the processor type. However, the method of arbitration is not defined in AMBA specifications. Instead, the arbiter is allowed to be designed as per the suitability of the application requirements. There are three distinct bus types specified within the AMBA bus. These are ASB, AHB and APB.

### 3.1.1 ASB or Advanced System Bus

It is meant to be used for simple cost-effective designs that support burst transfer, pipelined transfer operation, and multiple bus masters. This is the typical first generation bus of the AMBA system bus series.

### 3.1.2 AHB or Advanced High-performance Bus

This is a later generation bus of the AMBA bus system series. Unlike the ASB, it is intended for high performance designs and for providing communication channel with a high bandwidth. It supports multiple bus masters operation, peripheral and burst transfers, split transactions, wide data bus configurations, and non tristate implementations. Master, slave, arbiter, and decoder are the main constituents of the AHB. The high bandwidth communication channel is meant to serve between embedded processor (ARM, MIPS, AVR, DSP 320xx, 8051, etc.); and peripherals meant for high performance as well as hardware accelerators such as MPEG, ASICs, Colour LCD, etc; on-chip SRAM; APB bridge and on-chip external memory interface.

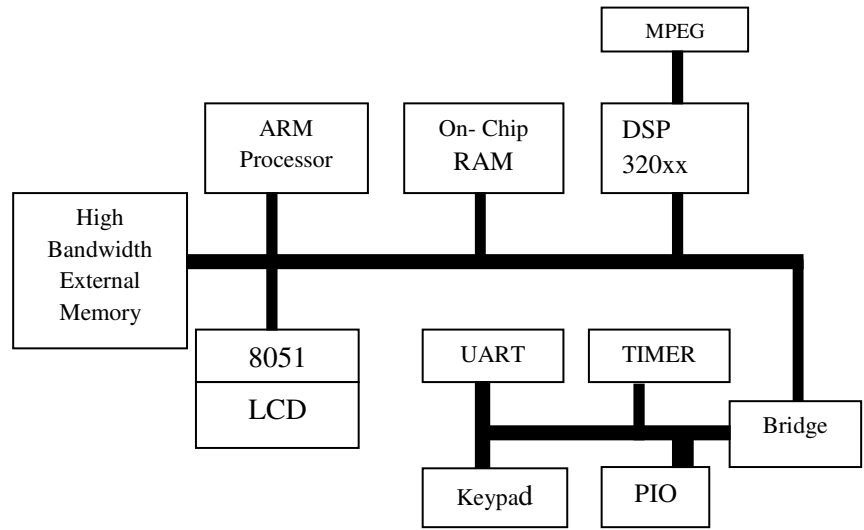

Figure – 5   AMBA based System Architecture

### 3.1.3 APB or Advanced Peripheral Bus

It is the bus system devised to connect general purpose low-speed low-power peripheral devices. Except for the bridge that is deployed as the peripheral bus master, all other APB bus devices such as the Timer, UART, PIA, etc. are used as slaves. In order to achieve the target of easy interfacing, the APB provides a simple addressing with latched addresses and control signals. This type of internal mechanism renders APB as a static bus. Recently, two new specifications (Multi-Layer AHB and AMBA AXI) have been added and defined under the AMBA bus system series. The Multi-layer AHB [20] provides more flexible interconnect architecture with respect to AMBA AHB. It is in the form of matrix which enables parallel access paths between multiple masters and slaves. It helps to keep the AHB protocol unchanged. On the other hand, AMBA AXI is based on the concept of point-to-point connection [21].

115

## 3.2 CoreConnect

CoreConnect, an on-chip bus developed by IBM [8], allows the reuse of processor, sub-system and peripheral core, supplied from different sources, and helps these to integrate into a single VLSI design. This bus architecture is hierarchically organized. "Fig. 6," shows not only the architecture of the three buses (PLB, OPB, and DCR) but also how these three combine to comprise the CoreConnect bus and its hierarchical organization. The architecture of this type helps in providing an efficient interconnection of cores, custom logic and library macros within a SoC

### 3.2.1    PLB or Processor Local Bus

This comprises the main system bus under CoreConnect. As per its features, it is a synchronous, multi-master, centrally arbitrated bus capable of achieving high-performance and low-latency on-chip communication. Concurrent read and write transfers are supported by separate address bus, and data bus. Interconnection of various master and slave macros is achieved by using PLB macro, as glue logic. PLB is attached to each PLB master through separate addresses, read or write data buses and control signals. On the other hand shared, but decoupled, read and writes data buses, addresses are used to attach the slaves to PLB. It can support up to 16 masters. However, there are no restrictions in the number of slave devices to be attached to the PLB [8].

### 3.2.2    OPB or On-chip Peripheral Bus

CoreConnect is designed to connect low speed, low throughput peripherals, (like serial and parallel port, UART, etc). As per its features, the OPB is capable of fully synchronous operation, dynamic bus sizing, separate address and data buses, multiple OPB bus masters. The bus cycle and data transfer features include single cycle transfer of data between OPB bus master and OPB slaves, single cycle transfer of data between bus masters, etc. In terms of its implementation the OPB is a multi-master, arbitrated bus that makes use of distributed multiplexer instead of tristate drivers. In case of the PLB masters trying to gain access to the peripherals on the OPB bus, it is allowed through the OPB bridge macro that acts as a slave device on the PLB and a master device on the OPB.

### 3.2.3    DCR bus or Device Control Register Bus

CoreConnect is a single master bus mainly used as an alternative (relatively low speed data path) to the system for two reasons. First, it is used for passing status and setting configuration information into the individual device-control- registers between the Processor Core and others SoC constituents such as Auxiliary Processors, On-Chip Memory, System Cores, Peripheral Cores etc. Second, it is used as a design for testability purposes. As per its features, DCR bus is a synchronous bus based on a ring topology which is implemented as distributed multiplexer across the chip. The address bus in DCR bus consists of 10-bits and the data bus consists of 32-bit. The arbitration in the CoreConnect is implemented on the basis of static priority, and with programmable priority fairness.

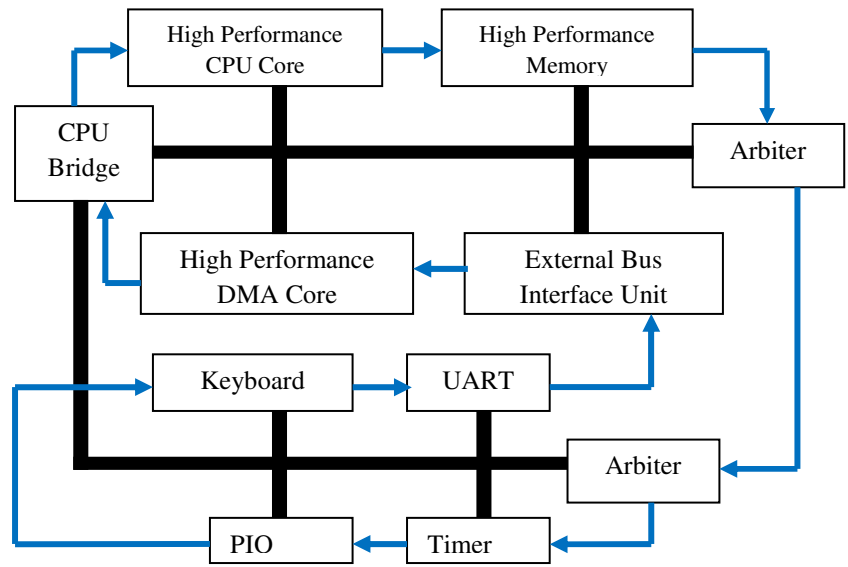

Figure – 6 CoreConnect Bus Architecture

## 3.3 AVALON

The Avalon bus [22] is the bus system owned and used by Altera Corporation to interconnect SOPC devices, especially the Nios processor(s) and other Avalon peripherals. Being an Altera's parameterized bus, Avalon is mainly used for FPGA SoC design based on Nios processor [23].
The Avalon implements simultaneous multi-master bus architecture. The main advantage of this architecture lies in its eliminating the bandwidth-bottleneck as it offers increased bandwidth between peripherals regardless of the bus standard that connects them. [24] In Avalon bus the bus masters contend for individual slaves, not for the bus itself, therefore, multiple masters can be active at the same time and can simultaneously transfer data to their slaves. As long as another master does not access the same slave at the same time,    Masters can access a target slave without any delay or waiting. As Arbitration is required when two masters contend for or connect to the same slave, Avalon implements distributed arbitration by using the technique called slave-side arbitration.

The Avalon bus supports an entirely in-built arbitration implementation for its Nios-based systems using this bus module. The Avalon bus is an active, on-chip bus architecture consisting of logic and routing resources inside *a PLD. It* supports a set of predefined signal types and a user can connect IP blocks with the help of the same. It uses separate address, data and control lines.
Avalon has a synchronous interface. It specifies the port connections between master and slave components. The Avalon specification also includes and indicates the timing by which the master and slave components communicate. Avalon includes a number of features and conventions to support automatic generation of systems, busses, and peripherals by the SOPC Builder software.

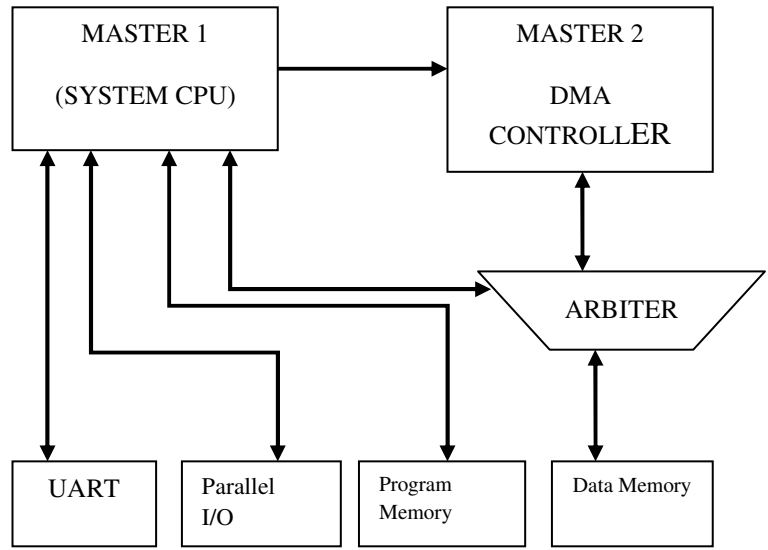

Figure – 7 Avalon Bus Architecture

In addition to the Multiple Masters Bus Architecture, the other features available to the Avalon supported peripherals include -- up to 4GBytes address space; synchronous interface; separate address, data and control lines; built-in address decoding; Wizard-based configuration; dynamic bus sizing*;* wait state generation; interrupt priority assignment; latent transfer capabilities; and a streaming Read and Write capabilities.

## *4.* COMPARISON OF SoC BUSES

It was Rudolf Usselmann of OpenCores, who attempted a comparison of noted significance on three bus types – AMBA, WISHBONE and CoreConnect [25]. Patrick Pelgrims, Tom Tierens and Dries Driessens have added Avalon bus with the above mentioned three bus types, thus carrying the number under analytical survey to four buses [26]. Another comparative survey has been made by Milica Miti´c and Mile Stojˇcev. It is based on a criterion comprising five parameters namely; Topology, Synchronicity, Arbitration, Bus width and Operating Frequency [27]. The present paper adds three more features -- Nature of the Open Source, Architecture and Data Transfer -- to the earlier existing parameters to make the comparison more comprehensive in its application in respect of four bus types briefly mentioned above in this part of the paper. A Comparative Table (i.e. Table -2) at the end sums up this section comparing the four bus types in their selective features.

### 4.1 Nature of the Open Source/ Standard

First three out of the four bus types included in this review are open standard buses. No license /royalty are required to develop and sell products that use, or are based on the interfaces of these buses. AMBA and CoreConnect are open and free but require a registration before use. The user is required to register though free of cost before using their interfaces. Avalon leans a bit towards proprietary nature. The ModelSim-Altera Starter Edition software of Avalon, the Quartus II Web Edition software of Avalon and Altera IP mega functions in Avalon [28], do not require license files, but one needs a valid license file to run and use other software manufactured and marketed by this company. However, among all the four buses discussed in this paper, it is the WISHBONE bus which is open standard in the real sense because it doesn't bind its user in any way. It is not copyrighted. It requires no license agreement or registration at all. As it is in the public domain, its IP cores are available free for use to every user. In other words, the WISHBONE standard may be used for designing and production of integrated circuit components without royalties or other financial obligations.

## 4.2 Architecture

In terms of architecture AMBA and CoreConnect support hierarchical structure [29]. There are three levels of hierarchy in the CoreConnect bus. These are: On-chip Peripheral Bus (OPB), Processor Local Bus (PLB) and Device Control Register (DCR) bus. AMBA (Advanced Micro controller Bus Architecture) from ARM has two levels of hierarchy. These are: the Advanced High performance Bus (AHB), and the Advanced Peripheral Bus (APB). While AHB is similar to PLB in CoreConnect, the APB of AMBA is similar to OPB in CoreConnect. However, WISHBONE and Avalon do not support hierarchical structure. WISHBONE, in the hierarchical view of its architecture only supports structured design methodologies [30]. All other three buses except WISHBONE are capable of supporting features of pipelined architecture. In WISHBONE the possibility of this type of feature may only occur under the Dataflow interfacing. All the four bus types are similar in terms of supporting the multiplexed structure of architecture. AMBA and Avalon are (Multi) MASTER- (Multi) SLAVE buses and their arbitration scheme depends upon logic interface. CoreConnect also deploys (Multi) MASTER- (Multi) SLAVE architecture but in its case the maximum limit of the MASTERS deployed is eight only. Its arbitration depends on different priority schemes available. The architecture of WISHBONE bus is also (Multi) MASTER / (Multi) SLAVE but its arbitration logic is user defined.

## 4.3 Topology

The topology refers to the way SoC components are connected. It can be in the form of single shared architecture, dedicated communication channels or more complex architectures such as hierarchical buses, token ring or crossbars. AMBA makes use of hierarchical bus topology; Avalon deploys the point to point topology; CoreConnect is different from these because its topology has the shared data lines while the control lines form a point to point ring; the topology of WISHBONE is comparatively open ended as it may make use of a point to point, a ring, a shared bus or a cross-bar interconnection network. The multi-master capability of WISHBONE enables it for multi-processing [31].

## 4.4 Synchronicity

Synchronicity is a feature of Clocking in a bus. If a single clock is used for the communication medium and its connected cores, the system is referred to as a synchronous system. Communication medium synchronization occurs with the help of handshaking protocol that uses request-acknowledgement signals to ensure that data transfer is completed successfully. Handshaking is also used in synchronous systems for a data transaction consisting of several data transfers. All the four buses included in this review are synchronous buses.

### 4.5 Arbitration

AMBA (except APB which doesn't require any arbitration) uses application specific arbitration; Avalon makes use of distributed arbitration also popular as the slave- side- arbitration; CoreConnect depends on programmable priority fairness arbitration while WISHBONE deploys application specific arbitration.

### 4.6 Bus Width

The Data bus width in AMBA is 8, 16, 32 bits for APB and 32, 64, 128 or 256 bits for AHB and ASB. The address bus width is 32 bits for AMBA. The Data bus width is 1 to 128 and the address bus width is 1 to 32 bits for Avalon. The Data bus width of CoreConnect is 32, 64, 128 or 256 bytes for PLB; 8, 16 or 32 bytes for OPB; 32 bytes for its DCR while the address bus width of CoreConnect is 32 bytes for PLB and OPB and 10 bytes for its DCR variant. In the case of WISHBONE, the Data bus width is 8, 16, 32, or 64 bits and the address bus width is 1 to 64 bits.

### 4.7 Operating Frequency

AMBA and WISHBONE are quite similar in terms of their Operating Frequency because both are user defined. In the case of CoreConnect the operating frequency depends on the PLB width. In the case of Avalon it is not applicable.

### 4.8 Data Transfer

AMBA and CoreConnect can transfer data through Handshaking, Pipelined, Split and Burst transfer. Avalon makes use of Pipelined and Burst transfer mode. WISHBONE deploys Handshaking Protocol and Burst Transfer mode for the specific purpose of data transfer.

Table - 2 Comparisons of WISHBONE Bus, AMBA Bus, Avalon Bus and CoreConnect Bus

| Sr. No. | Main Feature | Sub Features | Name of the Bus and its Owner | | | |
|---|---|---|---|---|---|---|
| | | | WISHBONE (1) | AMBA (2) | Avalon (3) | CoreConnect (4) |
| 1. | Originator/ Owner Of the Bus | ------ | OpenCores | ARM | Altera Corporation | IBM |
| 2. | Status | Open Architecture | Yes | Yes | Yes (Partial) | Yes |
| | | Registration | No | Yes | Yes | Yes |
| | | License | No | No | Yes | No |
| 3. | Architecture | Hierarchical | No | Yes | No | Yes |
| | | Pipelined | No | Yes | Yes | Yes |
| | | Multiplexed | Yes | Yes | Yes | Yes |
| 4. | Topology | Point-to-Point | Yes | No | Yes | No |
| | | Dataflow / Ring | Yes | No | No | Yes |
| | | Shared bus (unlevel) | Yes | No | No | No |
| | | Crossbar Switch | Yes | Yes | No | No |
| 5. | Arbitration | Static Priority | Yes (Application Specific) | Yes (Application Specific (Except APB) | Yes (Slave Side) | Yes |
| | | TDMA | Yes (Application Specific) | Yes (Application Specific) (Except APB) | Yes (Slave Side) | No |
| | | CDMA | Yes (Application Specific) | Yes (Application Specific) (Except APB) | Yes (Slave Side) | No |
| | Arbitration | Round Robin | Yes (Application Specific) | Yes (Application Specific) (Except APB) | Yes (Slave Side) | No |
| | | Lottery | Yes (Application Specific) | Yes (Application Specific) (Except | Yes (Slave Side) | No |
| | | Token Passing | Yes (Application Specific) | Yes (Application Specific) (Except APB) | Yes (Slave Side) | No |
| 6. | Data Transfer | Handshaking | Yes | Yes | No | Yes |
| | | Pipelined | No | Yes | Yes | Yes |
| | | Split Transfer | N/A | Yes | No | Yes |
| | | Burst Transfer | Yes | Yes | Yes | Yes |
| 7. | Bus Width | System bus or Data bus width (In Bits) | 8 – 64 | 8 -- 1024 | 1 – 128 | 64--2048 |
| | | Peripheral bus or Address bus width (in Bits) | 1 -- 64 | 1 – 32 | 1 -- 32 | 8 -- 256 |
| 8. | Clocking | Synchronous | Yes | Yes | Yes | Yes |
| | | Asynchronous | No | No | No | No |
| 9. | Operating Frequency | ------- | User defined | User defined | User defined | Depends on PLB Width |

## 5. CONCLUSION

A survey of the WISHBONE bus and its comparison with three other buses AMBA from the ARM, CoreConnect from the IBM and Avalon by the Altera Corporation reveals that in terms of compared performance parameters, the WISHBONE bus tends to gain an upper edge over the other three types because it provides for connecting circuit functions together in a way that is simple, flexible and portable [32] due to its synchronous design. It aides the system integrator by standardizing the IP Core interfaces that makes it much easier not only to connect the cores but also to create a custom System-on-Chip. The performance parameters like variable interconnection and variable timing specification provide flexibility to its programming process and its frequency range respectively. The WISHBONE bus differs from other buses over the issues of registration before use, in offering Support and Development Tools, in terms of designing one's own Libraries for plug-and-play logic utilization, in terms of Bus Architecture and Transfer Cycles because WISHBONE offers Read-Modify-Write (RMW) transfer that none of the other bus architectures does. At the end, this paper endorses the view held by Rudolf Usselmann [33] that it would be a wise choice to adopt WISHBONE as a primary interface to our cores because its signaling appears to be very intuitive and should be easily adopted to the other interfaces when needed.

International Journal of VLSI design & Communication Systems (VLSICS) Vol.3, No.2, April 2012

## Authors

Mohandeep Sharma received his B.Tech degree in Electronics and Communication From Shiv Shankar Institute of Engg. & Technology (SSIET) Patti affiliated to the Punjab Technical University, Jalandhar. Presently, he is pursuing his M.Tech degree in VLSI Design from the Centre for Development of Advanced Computing (CDAC), Mohali and is working on his thesis. His area of interest is VLSI Design, System on Chip, etc. His email address is mdsharma007@gmail.com.

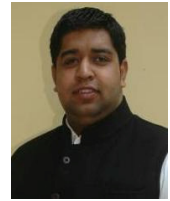

Dr. Dilip Kumar received his Ph.D. degree from MM University Mulana (Haryana). Currently, he heads the ACS Division at the Center for Development of Advanced Computing (CDAC), Mohali. During his 8 years of Teaching Experience, he has guided and supervised several project works as well as thesis works.The areas of his interest include Embedded System, Wireless Systems, VLSI Design and System on Chip. He may be contacted at his email address:    dilip.k78@gmail.com .

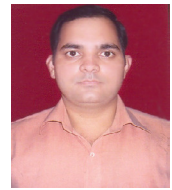